\documentclass{ifacconf}

\usepackage{graphicx}      
\usepackage{natbib}        
\usepackage{amsmath}
\usepackage{amssymb}
\usepackage{booktabs}
\usepackage{layouts}
\usepackage{xcolor}
\usepackage{algorithm}
\usepackage{pifont}
\usepackage[absolute,verbose]{textpos}
\newcommand{\cmark}{\ding{51}}
\newcommand{\xmark}{\ding{55}}
\usepackage{tikz}

\DeclareMathOperator*{\argmax}{arg\,max}
\DeclareMathOperator*{\argmin}{arg\,min}

\begin{document}
\begin{frontmatter}

\title{Robust Parametrization of a Model Predictive Controller for a CNC Machining Center Using Bayesian Optimization \thanksref{footnoteinfo}}

\thanks[footnoteinfo]{The presented research was funded by the Deutsche Forschungsgemeinschaft (DFG, German Research Foundation) under Germany's Excellence Strategy - EXC-2023 Internet of Production - 390621612.}

\author[First] {David Stenger} 
\author[First] {Muzaffer Ay} 
\author[First] {Dirk Abel}

\address[First]{Institute of Automatic Control, RWTH Aachen University, 
   Aachen, Germany (e-mail: \{d.stenger,m.ay,d.abel\}@irt.rwth-aachen.de).}

\begin{abstract}                
Control algorithms such as model predictive control (MPC) and state estimators rely on a number of different parameters. The performance of the closed loop usually depends on the correct setting of these parameters. Tuning is often done manually by experts based on a simulation model of the system. Two problems arise with this procedure. Firstly, experts need to be skilled and still may not be able to find the optimal parametrization. Secondly, the performance of the simulation model might not be able to be carried over to the real world application due to model inaccuracies within the simulation. With this contribution, we demonstrate on an industrial milling process how Bayesian optimization can automate the tuning process and help to solve the mentioned problems. Robust parametrization is ensured by perturbing the simulation with arbitrarily distributed model plant mismatches. The objective is to minimize the expected integral reference tracking error, guaranteeing acceptable worst case behavior while maintaining real-time capability. These verbal requirements are translated into a constrained stochastic mixed-integer black-box optimization problem. A two stage min-max-type Bayesian optimization procedure is developed and compared to benchmark algorithms in a simulation study of a CNC machining center. It is showcased how the empirical performance model obtained through Bayesian optimization can be used to analyze and visualize the results. Results indicate superior performance over the case where only the nominal model is used for controller synthesis. The optimized parametrization improves the initial hand-tuned parametrization notably.
\end{abstract}


\end{frontmatter}

\begin{textblock*}{\textwidth}(13.85mm,30mm)
	\small\textcopyright 2020 the authors. This work has been accepted to IFAC for publication under a Creative Commons Licence CC-BY-NC-ND.
\end{textblock*}

\section{Introduction}

A major goal of controller synthesis is to achieve best possible closed-loop performance. Besides choosing a suitable controller and possibly state estimator, their correct parametrization can heavily improve closed-loop performance. Analytical design rules are only available for a limited number of controllers and scenarios. Manual tuning of controller parameters is tedious and often suboptimal. Mathematical optimization can lead to a more systematic approach and overall better performance.

The problem of finding the optimal configuration of a controller can in the most simple case be written as:

\begin{equation} \label{eq:optproblem}
\boldsymbol{\theta}_c^{\text{opt}} = \argmin_{\boldsymbol{\theta}_c \in \mathcal{D}} C(\boldsymbol{\theta}_c) 
\end{equation}

The value of the loss function to be minimized $C(\boldsymbol{\theta}_c)$ is obtained for one set of controller parameters through simulation or experiment. This results in three challenges for the optimizer used to find the solution to Eq. \ref{eq:optproblem}: Firstly, we have a black-box optimization problem. Secondly, simulations and experiments might be time consuming requiring a sample efficient optimization algorithm. Thirdly, the objective function might be corrupted by noise $\tilde{C}(\boldsymbol{\theta}_c) = C(\boldsymbol{\theta}_c) + \epsilon$. These challenges rule out a number of common optimization algorithms. Algorithms relying on gradients or relaxations of the problem are not suitable. Meta-heuristics such as Genetic Algorithms or Particle Swarm Optimization are considered to be not sample efficient since they discard some of the previously obtained simulation results. Additionally, they do not incorporate stochastic constraints and objective functions inherently. 

Bayesian optimization is a common method used for noisy optimization. Surrogate models of black-box responses are learned by fitting fast-to-evaluate probabilistic regression models using all data obtained through previous sampling of the black-box during optimization. These surrogate models are used to find the next promising sample point.

Bayesian optimization is a tool widely used in algorithmic tuning across different domains such as tuning of optimization algorithms [\cite{Hutter.2011}] and Fault detection and isolation (FDI) [\cite{Marzat.2011}].   
Recent examples from the control and robotics community include learning gaits under uncertainty [\cite{Calandra.2016}], throttle valve control [\cite{NeumannBrosig.2019}], local linear dynamics learning [\cite{Bansal.2017}] and tuning for a linear quadratic regulator for robots [\cite{Marco.2016}]. 
A related line of research is safe Bayesian optimization which tries to prevent the algorithm from sampling in unsafe controller parameter regions and therefore allows experimental parameter tuning [\cite{Berkenkamp.2016}]. 
 This method was recently applied by \cite{Khosravi.2019}. 

In this contribution we apply Bayesian optimization to an industrial CNC machining process with machinery sensitive to improper actuation. We cannot expect this process to be modeled accurately with low computational cost a-priori\footnote{Accurate modeling is possible via dexel simulations. Computational costs prohibit their usage in controller synthesis.}. Therefore the focus of this contribution is to find a robust parameterization for the combination of a Kalman Filter (KF) and a MPC with respect to random model plant mismatches in the simulation. Goal is to maximize expected performance while ensuring safe worst case behavior.

Parameter tuning for MPC with Bayesian optimization was recently examined by \cite{Piga.2019} and \cite{Lucchini.2020}. However, the authors do not consider save worst case behavior or optimization of the state estimator. \cite{Andersson.2016} optimize tuning parameters of soft constraints for stochastic collision avoidance with MPC taking into account probabilistic safety constrains.

Ensuring minimal worst case behavior has been previously addressed in the FDI community by \cite{Marzat.2011}. \cite{Ghosh.2018} used Bayesian optimization to search for adversarial examples for controllers. \cite{Krause.2011} considered finding optimal parameters for different application-contexts using Bayesian optimization.

Our main contributions in the context of parameter tuning for MPC with Bayesian optimization are to 
\begin{itemize}
	\item ensure acceptable worst case behavior by explicitly considering model uncertainty while 
	\item simultaneously optimizing the MPC and KF \newline parametrization and 
	\item expand previous work on outlier detection by adding a classification step to prohibit resampling in an area where outliers often occur.
\end{itemize}

The presented method can be used to find a safe parametrization to initiate experimental manual tuning, tuning with safe Bayesian optimization [\cite{Berkenkamp.2016}] or other online learning methods.

The paper is organized as follows: In Section \ref{sec:PS}, the application system is described in detail and the optimization problem is stated. In Section \ref{sec:BayesOpt}, Bayesian optimization including design choices made for the problem at hand is introduced. Section \ref{sec:MinMaxBayesOpt} shows how it is used in a two stage framework to solve the problem described in Section \ref{sec:PS}. In Section \ref{sec:Results}, the presented approach is evaluated. 

\section{Problem Statement} \label{sec:PS}

 Milling is a fast and flexible machining process, which is highly acclaimed in production due to its productivity. During milling, a rotating tool is moved against a workpiece to cut material along a predefined trajectory. Thus, a desired geometry can be manufactured. The presented optimization approach is examined for the quality control during milling. The quality of the production is defined by the deviation of the manufactured geometry from the desired geometry. This deviation relates to the deflection of the working tool during the process. Hence, a reproducible quality for the milling process requires the control of the cutting force, which leads to tool deflection. The force in turn relates to the feed velocity of the tool. Therefore, a multi-stage approach is applied, where on the outer loop the trajectory for the feed velocity is optimized with respect to the resulting force and on the inner loop the feed velocity of the tool is controlled. The tool dynamic is described as
 
 \vspace{-0.2cm}\begin{equation}
 \ddot{v}\left(t\right) + 2 \, D \, \omega_0 \, \dot{v}\left(t\right) + \omega_0^2 \, v\left(t\right) = K \, \omega_0^2 \, u\left(t-t_d\right),
 \end{equation}
 
 where $v\left(t\right)$ stands for the tool velocity at time $t$ in dependence of the control input $u$ after the delay-time $t_d$ and the model parameters $K$ for gain, $D$ for damping and $\omega_0$ for resonance frequency. 
 
 In order to avoid high cutting forces before occurrence, a model predictive control strategy is applied for the quality control during milling. In addition, a Kalman-Filter (KF) is used for state-estimation during the process. The reader may refer to previous work by \cite{Stemmler.2019} for further details about the structure of the control approach. 

The performance of the overall control strategy depends in part on the parametrization of the MPC and KF. Namely for the MPC, the prediction horizon $H_\text{p}$, control horizon $H_\text{u}$ are relevant parameters. Additionally, the ratio $\lambda_{\text{MPC}} \in \mathbb{R}$ between the weighting matrices $\mathbf{Q} = \mathbf{I}$ and $\mathbf{R} = \mathbf{I} \, 10^{\lambda_{\text{MPC}}}$ in the MPC cost function,

\vspace{-0.2cm}\begin{equation}
	J = ||\boldsymbol{e}||_{\mathbf{Q}} + ||\boldsymbol{\Delta u}||_{\mathbf{R}}
\end{equation}

are considered within optimization. This way, the deviation from the desired trajectory $\boldsymbol{e}$ and the change of the control input $\boldsymbol{\Delta u}$ can be weighted differently during optimization. Regarding the KF, the covariance matrix for measurement noise is predefined as $\mathbf{R}_{KF} = \mathbf{I} \cdot 0.001$ which corresponds to an experimentally determined variance of the sensor. In order to weight between prediction and measurement at correction, the covariance matrix for process noise is set as $\mathbf{Q}_{\text{KF}} = \mathbf{I} \, 10^{\lambda_{\text{KF}}}$ by the ratio $\lambda_{\text{KF}} \in \mathbb{R}$.

Satisfactory tracking performance for the underlying control of the feed velocity was achieved with the hand-tuned default parametrization ($H_\text{u} = 15$, $H_\text{p} = 15$, $\lambda_{\text{MPC}} = -3$, $\lambda_{\text{KF}} = -1$) on the nominal model. In order to increase the robustness of the parameterization, sources of uncertainty are introduced in the simulation model. Measurement signals are perturbed with zero mean Gaussian noise. Additionally, a model plant mismatch is introduced by modifying the plant models stiffness $\tilde{\omega}_0 = \omega_0 \cdot \theta_{e,1}$ and damping  $\tilde{D} = D \cdot \theta_{e,2}$. Note that while the simulated process model is disturbed, the internal MPC model is kept constant. From now on we refer to certain realizations of $\boldsymbol{\theta}_{e} =\{\theta_{e,1}, \theta_{e,2}\}$ as context. Note that in general other types of environmental conditions can be included in the contextual variables such as different reference trajectories or operating modes.    

\begin{figure}[ht!] 
	\includegraphics[width=\linewidth]{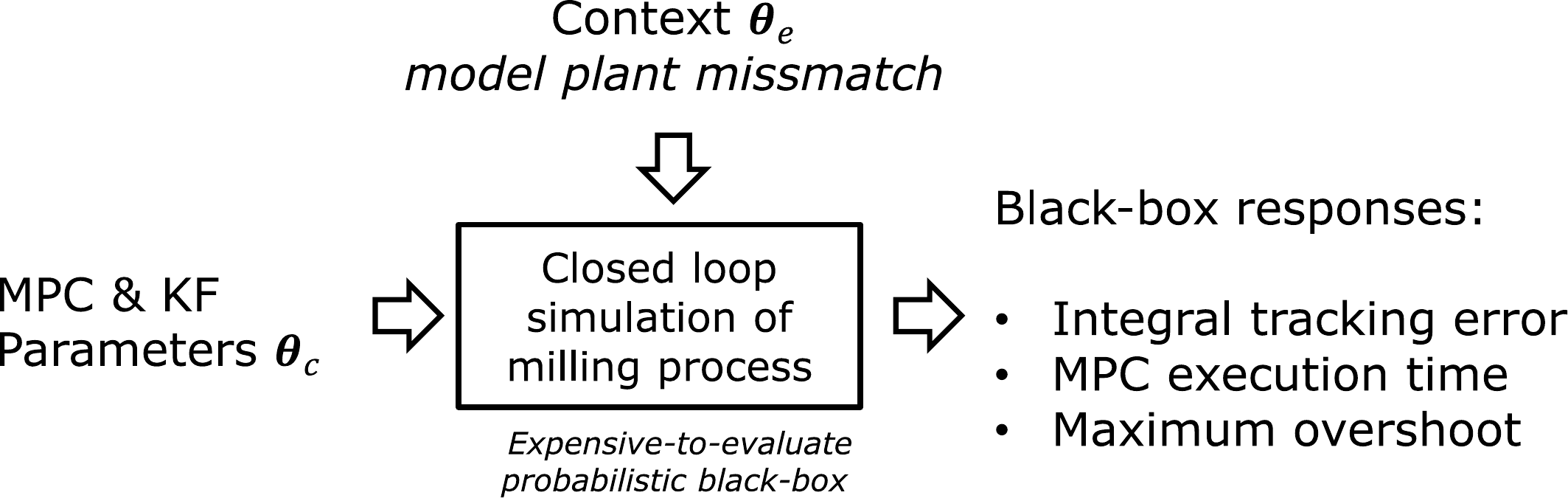}
	\caption{Overview of the simulation environment}
	\label{fig: Overview}
\end{figure}

Goal now is to find a robust set of parameters $\boldsymbol{\theta}_c = \{H_\text{u} , H_\text{p}, \lambda_{\text{MPC}}, \lambda_{\text{KF}} \}$ with respect to a known probability distribution of $\boldsymbol{\theta}_{e}$. The inputs and outputs of the simulation model used during optimization are shown in Fig. \ref{fig: Overview}. Note that even in a fixed context, the simulation is probabilistic because the realization of measurement noise is drawn at random.

The optimization problem is stated in Eq. \ref{eq:FullOptProblem}. The objective function (Eq. \ref{eq:2a}) is the expected (with respect to the assumed probability distribution of $\theta_{e}$) value of the integral tracking error, $C_{\text{EITE}}(\theta_{\text{c}})$. For one simulation it is calculated by comparing the actual velocity $v_k$ at timestep $k$ with the reference velocity $v_{\text{ref},k}$. Here it is assumed that $\boldsymbol{\theta}_{e}$ is distributed according to a truncated normal distribution with no correlation in between components of $\boldsymbol{\theta}_{\text{e}}$ as follows: 

\begin{equation}
\boldsymbol{\theta}_e \in \mathbb{R}^2 \sim  \mathcal{\psi}(\boldsymbol{\mu}_e,\sigma_e^2\mathbf{I},\boldsymbol{\theta}_e^{\text{min}},\boldsymbol{\theta}_e^{\text{max}}) \label{eq:2d} 
\end{equation}

where $\mathcal{\psi}(\boldsymbol{\mu}_e,\sigma_e^2\mathbf{I},\boldsymbol{\theta}_e^{\text{min}},\boldsymbol{\theta}_e^{\text{max}})$ is a normal distribution with $p(\boldsymbol{\theta}_e < \boldsymbol{\theta}_e^{\text{min}}) = p(\boldsymbol{\theta}_e > \boldsymbol{\theta}_e^{\text{max}}) = 0$.

Note however that the presented approach is in principle applicable for all probability distributions with bounded support. In order to prevent damage to the milling head the maximal overshoot in the worst case context $\Delta h^*(\boldsymbol{\theta}_c)$ is constrained (Eq. \ref{eq:2f}). This constraint is not explicitly considered in the original MPC formulation\footnote{This constraint can be integrated in the MPC formulation. This might increase calculation time and its fulfillment cannot be guaranteed if the MPC model is inaccurate as it is in the presented case.}. In addition, to be able to use cost efficient hardware the maximal computation time $T(\boldsymbol{\theta}_e|\boldsymbol{\theta}_c)$ is limited (Eq. \ref{eq:2e}). With a probability of $\Phi(z = 3) = 0.96$ it is not allowed to exceed a critical value of $T^{\text{max}}$.

\begin{subequations}\label{eq:FullOptProblem}
	\begin{align} 
	\min_{\boldsymbol{\theta}_c} \quad & C_{\text{EITE}}(\boldsymbol{\theta}_c) = \mathbb{E} \left[\sum_{k = 0}^{N} (v_{\text{ref},k}-v_k(\boldsymbol{\theta}_e|\boldsymbol{\theta}_c))^2 \right]  \label{eq:2a}\\[7pt]
	s.\,t.\quad \, \, & \boldsymbol{\theta}_c \in \mathbb{Z}^2 \times \mathbb{R}^{2} \label{eq:2b} \\[7pt]
	& \boldsymbol{\theta}_c^{\text{min}} \leq  \boldsymbol{\theta}_c \leq \boldsymbol{\theta}_c^{\text{max}}  \label{eq:2c} \\[7pt]	
	&  p(T(\boldsymbol{\theta}_e|\boldsymbol{\theta}_c) < T^{\text{max}}) > \Phi(z = 3) = 0.96  \label{eq:2e} \\[7pt]  
	&  \max_{\boldsymbol{\theta}_e} (\Delta h(\boldsymbol{\theta}_e|\boldsymbol{\theta}_c)) = \Delta h^*(\boldsymbol{\theta}_c)  < \Delta h^{\text{max}} \label{eq:2f}
	\end{align}
\end{subequations}

It should be noted that the optimization methodology presented is not limited to the constraints and objective function chosen here. Arbitrary constraints or performance matrices can be used. Preferably those which cannot be explicitly considered in the controller.

\section{Noisy constrained Bayesian optimization with outlier detection}
\label{sec:BayesOpt}

The aim of this Section is to explain the fundamentals of noisy Bayesian optimization with outlier detection as well as highlight the design choices made for the problem at hand. Two instances of the algorithm described in this section are used in a hierarchical approach as will be described in Section \ref{sec:MinMaxBayesOpt}. For notational simplicity, we now consider a simplified version of the optimization problem stated in Eq. \ref{eq:FullOptProblem}: 
\begin{subequations}\label{eq:SimpleOptProblem}
	\begin{align} 
	\min_{x} \quad & \mathbb{E} \left[y(x)\right]  \label{eq:4a}\\[3pt]
	s.\,t.\quad \, \, & \boldsymbol{x} \in \mathbb{R}^{m} \label{eq:4b} \\[3pt]
	&  p(g(x) < g^{\text{max}}) > \Phi(z)  \label{eq:4c} 
	\end{align}
\end{subequations}
The objective function $y(x)$ as well as the constrained black-box response $g(x)$ are corrupted by noise. Each time the Black-Box is evaluated with parameters $x_i$, corresponding noisy samples $y_i$ and $g_i$ are obtained. 

\begin{algorithm}[h] 
	1: Initial sampling of $X_{1}$, $Y_{1}$ and $G_{1}$:  \\ [3pt]
	2: \textbf{for} k = 1; 2; . . . ; \textbf{do} \\[3pt]
	3: \quad update probabilistic surrogate models using  \\
	\hspace*{6.5mm} $\tilde{X}_{k+1}$,$\tilde{Y}_{k+1}$ and $\tilde{G}_{k+1} $\\[3pt]
	4: \quad select $x_{k+1}$ by optimizing an acquisition function:\\ 
	\hspace*{6.5mm} $x_{k+1} = \argmax_x(\alpha(x))$\\[3pt]	
	5: \quad query objective function to obtain $y_{k+1}$ and $g_{k+1}$ \\[3pt]
	6: \quad augment data $X_{k+1} = \{ X_{k}, x_{k+1}\}$, \\
	\hspace*{6.5mm} $Y_{k+1} = \{ Y_{k},y_{k+1}\}, \, G_{k+1} = \{ G_{k},y_{k+1}\}$, \\[3pt]
	7: \quad $\tilde{X}_{k+1},\tilde{Y}_{k+1},\tilde{G}_{k+1} \leftarrow \textrm{OutDetect}(X_{k+1},Y_{k+1},G_{k+1}$) \\[7pt]
	8: \textbf{end for} 
	\label{Algo:BayesOpt}
	\caption{Bayesian optimization with outlier detection}
\end{algorithm}

Algorithm 1 shows the procedure of Bayesian optimization. A detailed introduction to Bayesian optimization is provided for example by \cite{Shahriari.2016}. The main idea is to use all sample points obtained so far ($X_{k} = [x_1,\dots ,x_k],Y_{k} = [y_1,\dots ,y_k],G_{k} = [g_1,\dots ,g_k]$) to construct fast-to-evaluate surrogate models of $y(x)$ and $g(x)$ at each iteration (Line 3) and use these models to search for the next promising sample point. This way the surrogate models are iteratively refined in promising regions.  

In this work Gaussian process regression (GPR) models are used as surrogate models.  A separate GPR model is built for each of the responses. For a detailed introduction on GPR the reader is referred to \cite{Rasmussen.2006}. GPR is a nonparametric regression and interpolation model which provides a probabilistic prediction of each  of the black-box responses for parametrization which have not been evaluated yet. The model is here defined by a constant a-priori mean, a squared exponential kernel with automated relevant detection and a homoscedastic Gaussian observation model (to account for the noisy samples). Hyperparameters are optimized at each iteration by maximizing the likelihood. Hyper-priors are placed on the hyperparameters to include expert knowledge in the optimization and avoid potential over fitting. E.g. the lower bound on the kernel length scales in the direction of prediction and control horizon are set to $l_{H_\text{u}}^\text{min} = l_{H_\text{p}}^\text{min} = 0.22$ which roughly speaking corresponds to a covariance of the objective function values of at least $0.1$ if the horizons are changed by one. 

Based on these models, an acquisition function $\alpha(x)$ is used to balance between exploitation (sampling close to the current optimum) and exploration (sampling where the model is uncertain) when searching for the next sample point (Line 4). Here the so called reinterpolation procedure (RI) proposed by \cite{Forrester.2006} is used. Using RI prohibits multiple evaluations of the objective function
	for identical parameters by fitting an intermediate interpolating surrogate model. This is beneficial in the presented
	case since repeating the second stage of the optimization
	(cf. Section 4.2) twice for identical parameters would be
	a waste of computational resources. In addition to maximize the noisy expected improvement of the objective function by maximizing $RI(x_q)$, we want our query point to be feasible $p_{\text{feas}}(x_q)$, to not be an outlier $p_{\text{out}}(x_q)$ with off-the-charts objective function value and we want the simulation not to fail $p_{\text{fail}}(x_q)$. Therefore, the following acquisition function is used:   
 
\begin{equation}
\alpha(x_q) = RI(x_q)p_{\text{feas}}(x_q)(1-p_{\text{out}}(x_q))(1-p_{\text{fail}}(x_q)) \label{eq:InfillCriteria}
\end{equation}

The probability of feasibility is calculated by using the probabilistic predictions of $g(x)$. At each iteration, the next sample point is chosen by maximizing $\alpha(x_q)$ using particle swarm optimization. 

After evaluation of the new sample point (Line 5), outlier detection is performed (Line 7). In Bayesian optimization with GPR, special care needs to be taken of outliers. At the problem at hand, some parametrization may lead to very large objective function values. This is problematic because outliers can severely corrupt the probabilistic surrogate models, leading to unrealistically small length scales. In order to detect outliers we follow the approach presented by \cite{MartinezCantin.2017}. In this paper outliers are detected by first fitting a robust GPR model with a student-t observation model and then detecting the observations with a low likelihood. 

In the original paper, outliers are discarded from the set of observations.  We take this approach one step further by building a k-nearest-neighbor (knn) classifier\footnote{A squared exponential	kernel as distance metric and a squared inverse	distance weighting is used .} to estimate the probability $p_{\text{out}}(x_q)$ of observing an outlier at a given location $x_q$. The training data set of the knn-classifier consists of all sample points within the controller design parameter space labeled according to whether they were identified as outliers or not. 

For some parameterizations the control algorithm to be configured may lead the simulation environment to crash due to numerical issues. To prevent repetitive sampling in these areas an additional knn-classifier is built to estimate the probability $p_{\text{fail}}(x_q)$ of a parametrization $x_q$ to lead to a crash.

\section{Two stage Bayesian optimization for MPC and KF tuning}
\label{sec:MinMaxBayesOpt}

\begin{figure}[htb!] 
	\includegraphics[width=\linewidth]{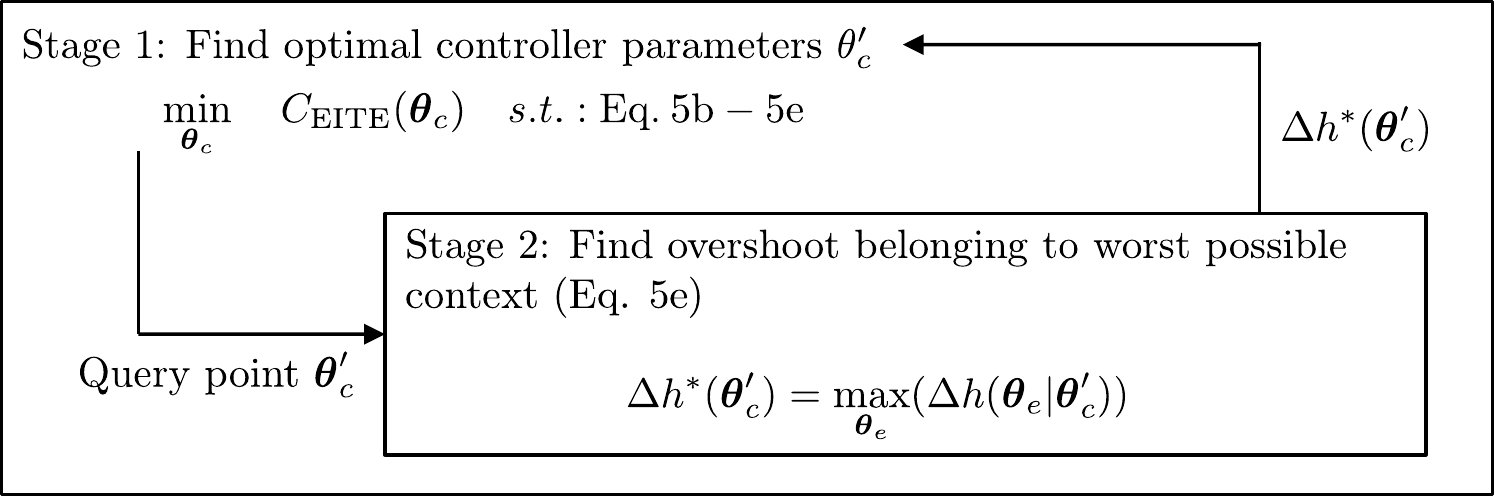}
	\caption{Overview of the two stage optimization approach}
	\label{fig: TwoStageOpt}
\end{figure}

 The algorithm presented in Section \ref{sec:BayesOpt} is applied within a two stage optimization approach to solve the problem stated in Eq. \ref{eq:FullOptProblem}. An overview is given in Fig. \ref{fig: TwoStageOpt}. Stage one has the task to find the optimal controller configuration characterized by low expected integral tracking error, acceptable worst case overshoot and real time capable execution time. For each controller configuration, $\boldsymbol{\theta}_{c}'$, queried by Stage one, the overshoot belonging to the worst possible combination of environmental variables, $\Delta h^*(\boldsymbol{\theta}'_c) = \max_{\boldsymbol{\theta}_e} (\Delta h(\boldsymbol{\theta}_e|\boldsymbol{\theta}'_c))$, is searched for in Stage two. This is achieved by solving a second stage optimization problem and ensures that constraint \ref{eq:2f} is satisfied. This procedure is similar to the one presented by \cite{Marzat.2011} in the context of FDI. 

\subsection{Stage one: Optimize controller parameters}

The algorithm presented in Section \ref{sec:BayesOpt} is used within Stage one. Although $p(\boldsymbol{\theta}_e)$ is assumed to be a truncated normal distribution, we cannot make any assumptions about the true structure of the probability density function of a response (i.e. integral tracking error, return time, and maximum overshoot) for a given controller parametrization $\boldsymbol{\theta}_c$ because the simulation model is non-linear. By using GPR with Gaussian likelihood (observation model) we approximate the true unknown probability distribution of the responses, by a deterministic function corrupted by Gaussian noise. This way in total 3 GPR models are built - one for each of the relevant black-box responses: $C_{\text{EITE}}(\boldsymbol{\theta}_c)$, $T(\boldsymbol{\theta}_c)$ and $\Delta h^*(\boldsymbol{\theta}_c)$. Latin hypercube sampling is used as initial sampling. As the objective function (Step 5 of Algorithm 1) of stage one, stage two of the optimization procedure is called.

\subsection{Stage two: Find worst possible context}\label{sec:StageTwo}

The goal of Stage two is to find the maximum overshoot for a given controller parametrization $\boldsymbol{\theta}_{c}'$ queried by Stage one. Note that using RI in the first stage prohibits multiple evaluations of the second stage with identical controller parameters.
 Noisy Bayesian optimization is used in this Stage, as well. But in this case only box constraints need to be considered. Outlier detection is also not used. Therefore, only one GPR model to approximate $\Delta h(\boldsymbol{\theta}_e|\boldsymbol{\theta}'_c)$ is built and standard RI is used as the acquisition function. Initial sampling is performed by drawing 5 different contexts from the truncated normal distribution of environmental variables.  
The corresponding 5 different integral tracking errors and return times of the initial sampling are returned to Stage one. The overshoot is maximized in the successive steps. Stage two is terminated either if an overshoot exceeding the maximum overshoot allowed was observed or if 10 evaluations were performed. Note that the GPR models built in both stages are completely independent from one another. This follows the approach by \cite{Marzat.2011}. An alternative approach would be to use a joint model using the controller parameter as well as the environmental parameters as explanatory variables for the overshoot as for example explained by \cite{Krause.2011}.

\section{Results} \label{sec:Results}

The algorithm presented in Sections \ref{sec:BayesOpt} and \ref{sec:MinMaxBayesOpt} is applied to the optimization problem stated in Section \ref{sec:PS} and was implemented in MATLAB 2017b using the GPML toolbox [\cite{Rasmussen.2010}] to create the GPR surrogate models. Additionally, a
simplified benchmark test case (Section \ref{sec:benchmark}) is used to compare the developed approach with three different benchmark approaches (Section \ref{sec:benchmarkAlgos}). Parameters are optimized with respect to a single reference velocity trajectory. This is reasonable because in a real world milling application we can expect to know the desired trajectory in advance. However, the presented approach is not limited to use only one single reference trajectory during optimization.

\subsection{Benchmark algorithms} \label{sec:benchmarkAlgos}
The presented algorithm (Algo. IV) is compared with three benchmark algorithms (Algo. I -III):

\begin{itemize}
	\item[I]    Bayesian optimization on nominal model
	\item[II]	Random sampling in Stage one
	\item[III]  Bayesian optimization without outlier detection
	\item[IV]   Bayesian optimization with outlier detection
\end{itemize}

The goal is to evaluate the performance impact of the individual proposed optimization steps. Algo. I is used for parameter optimization on the nominal model whereas Algo. (II-IV) are run on randomly varying realizations of the perturbed model using the second Stage presented in Section \ref{sec:StageTwo} to determine feasibility w.r.t. constraint \ref{eq:2f}. Algo. II uses random sampling in Stage one instead of finding the next sample point by maximizing the acquisition function presented in Eq. \ref{eq:InfillCriteria}. Algo. III is identical to the presented approach except that the outlier detection and classification scheme is not used.

\subsection{Benchmark test case} \label{sec:benchmark}
  
\begin{figure}[ht] 
	\includegraphics[width=\linewidth]{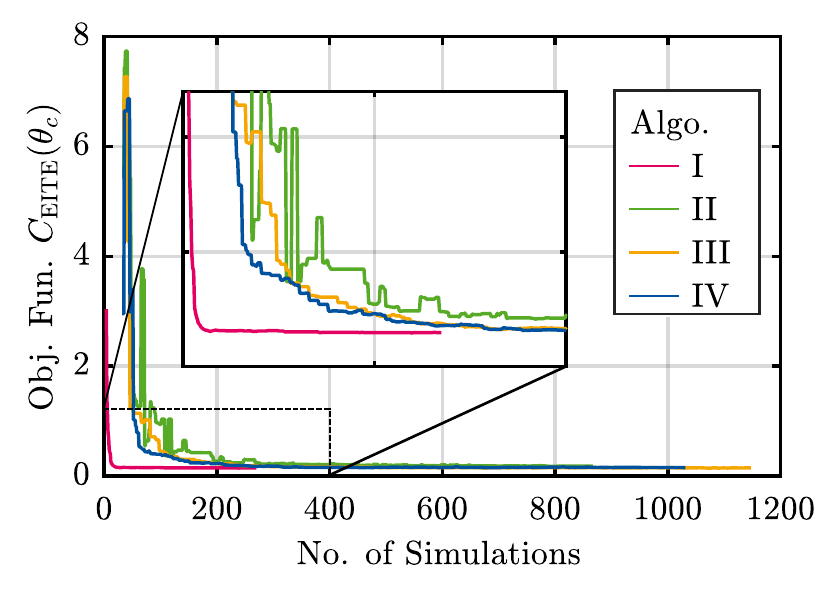}
	\caption{Validation performance after a given number of iterations considering only feasible solutions averaged over 10 runs (benchmark test case)}
	\label{fig: Performance}
\end{figure}

In order to compare the algorithms quantitatively each algorithm is run 10 times for 3 hours (only one hour for (Algo. I)) respectively on a simplified benchmark test case\footnote{All experiments were performed on a desktop PC with an AMD Ryzen 7 1700 Eight-Core Processor @3GHZ and 16 GB Ram}. The benchmark test case considers only two controller parameters $H_\text{u}$ and $\lambda_{\text{MPC}}$. The prediction horizon $H_\text{p}$ is set to the controller horizon $H_\text{u}$ and $\lambda_{\text{KF}}$ is kept at its hand-tuned default value. In each of the 10 runs, initial sampling was kept identical for each algorithm. The validation performance of a given parametrization is estimated by running the simulation with 25 different random draws from the distributions of model uncertainties. Table \ref{tab:Results1D} shows the average quality of the final solution of each of the algorithms. Fig. \ref{fig: Performance} depicts the average validation objective function of the feasible best solution after a given number of simulations during optimization.

\begin{table}[h] 
	\caption{Quality of the final solution averaged over 10 runs (benchmark test case)}
	
	\centering
	\begin{tabular}{ l r r r r  }
		\toprule
		
		Algorithm:                       &   I          & II                & III                & IV          \\
		\midrule
		Feasibility                &   $50\%$     &  $100\%$           &  $90\%$            & $100\%$ 	   \\ 
		Obj. fun. validation             &  $0.147$ 	& $0.182$           &  $0.148$           & $0.155$     \\ 
		$\Delta$ Obj. fun. val. - train  &   $+0.042$   & $+0.001$         &  $+0.003$          & $+0.007$    \\
	
		\bottomrule
	\end{tabular}
	\label{tab:Results1D}
\end{table}

It becomes apparent that only optimizing the MPC on the nominal model (Algo. I.) is not sufficient. Although it requires the algorithm only $\sim25$ simulations to converge, half of the time, the final solution is infeasible due to too much overshoot (Constraint \ref{eq:2f}) when model uncertainty is incorporated during validation.
Additionally, the expected integral tracking error is underestimated substantially. 

In contrast, Algos. II-IV are able to consistently find solutions which are feasible during validation and only slightly underestimate the validation integral tracking error. Random Sampling (Algo. II) is not able to find a competitive solution. Although the outlier detection and classification scheme used in Algo. IV improves the initial convergence (until around $200$ simulations), the average performance of the final solution is slightly better in Algo. III. 

Preliminary experiments have shown that the surrogate model of the objective function generated with outlier detection is more plausible than without outlier detection\footnote{Plots are not shown here due to space limitations.}. Without outlier detection, negative integral tracking errors are predicted in some regions and the surrogate models are far less smooth and in general uncertainty is larger. 

Furthermore, it was observed that with outlier detection, the hyper parameter optimization favored larger length scales which is consistent with the smoother predictions and smaller uncertainty. We hypothesize that the larger length scales hinder exploration in the later stages of optimization due to lower model uncertainty. Therefore interestingly, the more realistic fit obtained by GPR with outlier detection does not automatically mean that optimization performance is increased.

The experiments conducted by \cite{MartinezCantin.2017} have shown that outlier detection consistently improves the optimization procedure. This somewhat contradicts our findings. One reason is that in comparison to \cite{MartinezCantin.2017} we have \textit{deterministic} outliers, which consistently occur in one area of the design parameter space instead of \textit{random} outliers.

\begin{figure}     [htb!] 
	\vspace{0.1cm}
	\includegraphics[width=0.9\linewidth]{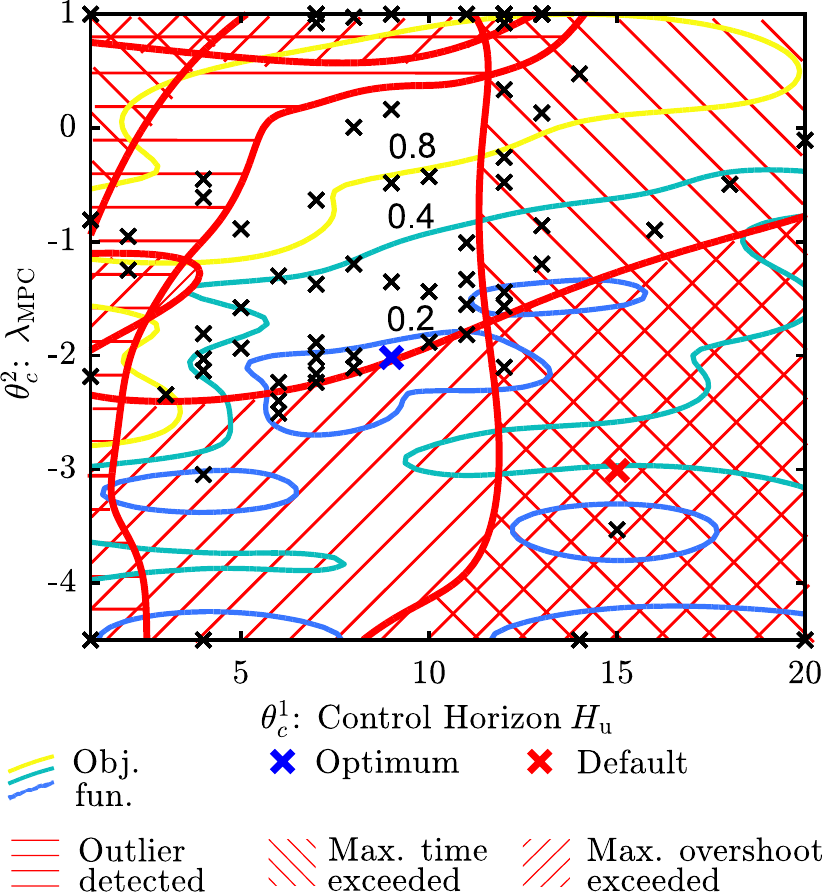}
	\centering
	\caption{Final surrogate model of the benchmark test case with estimated feasible area and objective function.} 
	\label{fig: FinalContour}
\end{figure}

Fig. \ref{fig: FinalContour} depicts the final estimated behavior of the closed loop within the controller design parameter space. In general the results are comprehensible. Calculation time is solely a function of $H_\text{u}$. Due to the presented outlier classification scheme we can observe that deterministic outliers (very large integral tracking errors) appear in areas with small $H_\text{u}$ and large $\lambda_{\text{MPC}}$. In contrast, the worst case overshoot constraint is violated for small values of $\lambda_{\text{MPC}}$ (i.e. low penalty on change of inputs). 

Interestingly, the optimum has a smaller $H_\text{u}$ than the calculation time constraint would allow. At first glance this may be counterintuitive. But given enough model uncertainty a longer prediction horizon does increase the chance of misprediction and therefore of critical overshoot. The default parametrization is located in the infeasible region. It was therefore considerably improved by the optimization approach.

\subsection{Full test case}

In the full test case all four parameters are considered for optimization $\boldsymbol{\theta}_{c} = \{H_\text{u},H_\text{p},\lambda_{\text{MPC}},\lambda_{\text{KF}}\}$. 
Algos. III \& IV were run 4 times for $6$ hours each.

The best and average validation performance as well as the average feasibility of the final solution are shown in Table \ref{tab:Results4Params}. 
Results indicate that the proposed algorithms can find a competitive solution even in a higher dimensional controller parameter space. Yet performances of the algorithms are less consistent than in the benchmark test case. Additionally, we can observe that the default parametrization of $\lambda_{KF} = -1$ and $H_\text{p} = H_\text{u}$ could not be improved upon.

\begin{table}[h] 
	\caption{Validation obj. fun. of the final solution averaged over 4 runs (full test case)}
	\centering
	\begin{tabular}{ r r r r}
		\toprule
		
		  					&   Best          & Average                & Feasibility   \\
		\midrule
		Algo. III           &   $0.143$                   &  $0.151$           					 &  $75\%$    \\
		Algo. IV            &   $0.141$                   &  $0.174$           					 &  $75\%$    \\	
		\bottomrule
	\end{tabular}
	\label{tab:Results4Params}
\end{table}

Temporarily, an alternative parametrization was considered to be the best feasible solution during the course of optimization. This parametrization is shown with some of its neighboring parameterizations in Table \ref{tab:ResultslocalMinimum}. Similar parameterizations were found in $50\%$ of the runs of Algos. III and IV. 

\begin{table}[h] 
	\caption{Parametrization with the best validation performance and neighboring parameterizations. }
	\centering
	\begin{tabular}{ r r r r | r c  }
		\toprule
		
		$H_\text{u}$   &   $H_\text{p}$          & $\lambda_{\text{MPC}}$                & $\lambda_{\text{KF}}$        & Obj. Fun.  & Feasibility          \\
		\midrule
		$\boldsymbol{1}$            &   $\boldsymbol{4}$                   &  $\boldsymbol{-3.9}$           					 &  $\boldsymbol{2}$                         & $\boldsymbol{0.12}$ 	 & \cmark  \\ 
		$1$            &   $3$                   &  $-3.9$           					 &  $2$                         & $0.19$ 	 & \textbf{\xmark}  \\
		$1$            &   $5$                   &  $-3.9$           					 &  $2$                         & $0.22$ 	 & \cmark  \\
		$1$            &   $4$                   &  $-3.9$           					 &  $-1$                        & $0.23$ 	 & \cmark  \\
		$2$            &   $4$                   &  $-3.9$           					 &  $2$                         & $0.14$ 	 & \xmark  \\
		
		\bottomrule
	\end{tabular}
	\label{tab:ResultslocalMinimum}
\end{table}

Although this parametrization shows superior validation performance in comparison to the best solution of the simplified test case, its neighboring solutions are either infeasible or the expected integral tracking error is unacceptably high.  
Therefore from a control engineering perspective this solution cannot be considered robust and would be rejected in practice. 

Fortunately, during later stages of the optimization these parameterizations are discarded by the optimization algorithms, because of the worse performing or infeasible neighborhood. This can be explained by the fact that in Bayesian optimization with GPR, black-box responses are assumed to be smooth on the characteristic kernel length scales. By setting a minimal kernel length scale based on domain knowledge as done in the present work, the optimizer implicitly considers the neighborhood of the optimal solution. This can be seen as an additional advantage of Bayesian optimization over other optimizers such as genetic algorithms where only the best solution is considered regardless of its neighborhood. Additionally, this alternative parametrization highlights the non-convexity of the optimization problem and of the relevance of all considered parameters.

\section{Summary \& Conclusion}

In this contribution Bayesian optimization was used to simultaneously optimize hyperparameters of a MPC and KF for an industrial CNC machining process. In order to achieve a robust parameterization, the simulation model was perturbed with model-plant mismatches randomly drawn from a known distribution with bounded support as well as randomly drawn measurement noise. Goal was to minimize expected integral tracking error, ensure worst case safety by constraining the maximum overshoot and enforce real-time capability by limiting the return time. The optimization problem is solved using a two-stage Bayesian optimization procedure relying on GPR with Gaussian observation model, the RI-acquisition function as well as outlier detection and classification. On a simplified benchmark test case, it was shown that optimization on the nominal model does not produce satisfactory parameter combinations. The default parametrization as well as random sampling was outperformed considerably. It was also observed that outlier detection did not consistently improve the convergence although surrogate models are more comprehensible. The empirical performance model obtained through Bayesian optimization allowed to analyze and visualize the results. Optimization on the full test case revealed the relevance of all parameters and the non-convexity of the optimization problem. Furthermore, it was shown how the model assumptions encoded in the hyperprior helps the optimization to avoid narrow and physically incomprehensible local minima. 

The presented approach can help control engineers to find an \textit{initial} robust and safe parametrization for controllers and state estimators given a closed loop simulation of the system and a probabilistic assumption over the model plant mismatch. It can empirically enforce constraints or performance metrics which are not or cannot explicitly be considered within the controller. This parametrization can then be further improved online for example by safe Bayesian optimization.

\bibliography{ifacconf}             

\begin{thebibliography}{19}
\providecommand{\natexlab}[1]{#1}
\providecommand{\url}[1]{\texttt{#1}}
\providecommand{\urlprefix}{URL }
\expandafter\ifx\csname urlstyle\endcsname\relax
  \providecommand{\doi}[1]{doi:\discretionary{}{}{}#1}\else
  \providecommand{\doi}{doi:\discretionary{}{}{}\begingroup
  \urlstyle{rm}\Url}\fi

\bibitem[{Andersson et~al.(2016)Andersson, Wzorek, Rudol, and
  Doherty}]{Andersson.2016}
Andersson, O., Wzorek, M., Rudol, P., and Doherty, P. (2016).
\newblock Model-predictive control with stochastic collision avoidance using
  bayesian policy optimization.
\newblock In \emph{2016 IEEE International Conference on Robotics and
  Automation (ICRA)}, 4597--4604. IEEE.
\newblock \doi{10.1109/ICRA.2016.7487661}.

\bibitem[{Bansal et~al.(2017)Bansal, Calandra, Xiao, Levine, and
  Tomiin}]{Bansal.2017}
Bansal, S., Calandra, R., Xiao, T., Levine, S., and Tomiin, C.J. (2017).
\newblock Goal-driven dynamics learning via bayesian optimization.
\newblock In \emph{2017 IEEE 56th Annual Conference on Decision and Control
  (CDC)}, 5168--5173. IEEE.
\newblock \doi{10.1109/CDC.2017.8264425}.

\bibitem[{Berkenkamp et~al.(2016)Berkenkamp, Schoellig, and
  Krause}]{Berkenkamp.2016}
Berkenkamp, F., Schoellig, A.P., and Krause, A. (2016).
\newblock Safe controller optimization for quadrotors with gaussian processes.
\newblock In \emph{2016 IEEE International Conference on Robotics and
  Automation (ICRA)}, 491--496. IEEE.
\newblock \doi{10.1109/ICRA.2016.7487170}.

\bibitem[{Calandra et~al.(2016)Calandra, Seyfarth, Peters, and
  Deisenroth}]{Calandra.2016}
Calandra, R., Seyfarth, A., Peters, J., and Deisenroth, M.P. (2016).
\newblock Bayesian optimization for learning gaits under uncertainty.
\newblock \emph{Annals of Mathematics and Artificial Intelligence}, 76(1-2),
  5--23.
\newblock \doi{10.1007/s10472-015-9463-9}.

\bibitem[{Forrester et~al.(2006)Forrester, Keane, and
  Bressloff}]{Forrester.2006}
Forrester, A., Keane, A., and Bressloff, N. (2006).
\newblock Design and analysis of noisy computer experiments.
\newblock \emph{AIAA Journal}, 44(10), 2331--2339.
\newblock \doi{10.2514/1.20068}.

\bibitem[{Ghosh et~al.(2018)Ghosh, Berkenkamp, Ranade, Qadeer, and
  Kapoor}]{Ghosh.2018}
Ghosh, S., Berkenkamp, F., Ranade, G., Qadeer, S., and Kapoor, A. (2018).
\newblock Verifying controllers against adversarial examples with bayesian
  optimization.
\newblock In \emph{2018 IEEE International Conference on Robotics and
  Automation (ICRA)}, 7306--7313. IEEE.
\newblock \doi{10.1109/ICRA.2018.8460635}.

\bibitem[{Hutter et~al.(2011)Hutter, Hoos, and Leyton-Brown}]{Hutter.2011}
Hutter, F., Hoos, H.H., and Leyton-Brown, K. (2011).
\newblock Sequential model-based optimization for general algorithm
  configuration.
\newblock In C.A.C. Coello (ed.), \emph{Learning and Intelligent Optimization},
  507--523. {Springer Berlin Heidelberg}, Berlin, Heidelberg.

\bibitem[{Khosravi et~al.(2019)Khosravi, Eichler, Schmid, Smith, and
  Heer}]{Khosravi.2019}
Khosravi, M., Eichler, A., Schmid, N., Smith, R.S., and Heer, P. (2019).
\newblock Controller tuning by bayesian optimization an application to a heat
  pump.
\newblock In \emph{2019 18th European Control Conference (ECC)}, 1467--1472.
  IEEE.
\newblock \doi{10.23919/ECC.2019.8795801}.

\bibitem[{Krause and Ong(2011)}]{Krause.2011}
Krause, A. and Ong, C.S. (2011).
\newblock Contextual gaussian process bandit optimization.
\newblock In {J. Shawe-Taylor}, {R. S. Zemel}, {P. L. Bartlett}, {F. Pereira},
  and {K. Q. Weinberger} (eds.), \emph{Advances in Neural Information
  Processing Systems 24}, 2447--2455. {Curran Associates, Inc}.

\bibitem[{Lucchini et~al.(2020)Lucchini, Formentin, Corno, Piga, and
  Savaresi}]{Lucchini.2020}
Lucchini, A., Formentin, S., Corno, M., Piga, D., and Savaresi, S.M. (2020).
\newblock Torque vectoring for high-performance electric vehicles: an efficient
  mpc calibration.
\newblock \emph{IEEE Control Systems Letters}, 1.
\newblock \doi{10.1109/LCSYS.2020.2981895}.

\bibitem[{Marco et~al.(2016)Marco, Hennig, Bohg, Schaal, and
  Trimpe}]{Marco.2016}
Marco, A., Hennig, P., Bohg, J., Schaal, S., and Trimpe, S. (2016).
\newblock Automatic lqr tuning based on gaussian process global optimization.
\newblock 270--277.
\newblock \doi{10.1109/ICRA.2016.7487144}.

\bibitem[{Martinez-Cantin et~al.(2018)Martinez-Cantin, Tee, and
  McCourt}]{MartinezCantin.2017}
Martinez-Cantin, R., Tee, K., and McCourt, M. (2018).
\newblock Practical bayesian optimization in the presence of outliers.
\newblock In A.~Storkey and F.~Perez-Cruz (eds.), \emph{Proceedings of the
  Twenty-First International Conference on Artificial Intelligence and
  Statistics}, volume~84 of \emph{Proceedings of Machine Learning Research},
  1722--1731. PMLR, Playa Blanca, Lanzarote, Canary Islands.

\bibitem[{Marzat et~al.(2011)Marzat, Piet-Lahanier, and Walter}]{Marzat.2011}
Marzat, J., Piet-Lahanier, H., and Walter, E. (2011).
\newblock Min-max hyperparameter tuning, with application to fault detection.
\newblock \emph{18th IFAC World Congress}.
\newblock \doi{10.3182/20110828-6-IT-1002.00476}.

\bibitem[{Neumann-Brosig et~al.(2019)Neumann-Brosig, Marco, Schwarzmann, and
  Trimpe}]{NeumannBrosig.2019}
Neumann-Brosig, M., Marco, A., Schwarzmann, D., and Trimpe, S. (2019).
\newblock Data-efficient auto-tuning with bayesian optimization: An industrial
  control study.
\newblock \emph{IEEE Transactions on Control Systems Technology}, 1--11.
\newblock \doi{10.1109/TCST.2018.2886159}.

\bibitem[{Piga et~al.(2019)Piga, Forgione, Formentin, and Bemporad}]{Piga.2019}
Piga, D., Forgione, M., Formentin, S., and Bemporad, A. (2019).
\newblock Performance-oriented model learning for data-driven mpc design.
\newblock \emph{IEEE Control Systems Letters}, 3(3), 577--582.
\newblock \doi{10.1109/LCSYS.2019.2913347}.

\bibitem[{Rasmussen and Nickisch(2010)}]{Rasmussen.2010}
Rasmussen, C.E. and Nickisch, H. (2010).
\newblock Gaussian processes for machine learning (gpml) toolbox.
\newblock \emph{J. Mach. Learn. Res.}, 11, 3011--3015.

\bibitem[{Rasmussen and Williams(2006)}]{Rasmussen.2006}
Rasmussen, C.E. and Williams, C.K.I. (2006).
\newblock \emph{Gaussian Processes for Machine Learning (Adaptive Computation
  and Machine Learning)}.
\newblock The MIT Press.

\bibitem[{Shahriari et~al.(2016)Shahriari, Swersky, Wang, Adams, and
  de~Freitas}]{Shahriari.2016}
Shahriari, B., Swersky, K., Wang, Z., Adams, R.P., and de~Freitas, N. (2016).
\newblock Taking the human out of the loop: A review of bayesian optimization.
\newblock \emph{Proceedings of the IEEE}, 104(1), 148--175.
\newblock \doi{10.1109/JPROC.2015.2494218}.

\bibitem[{Stemmler et~al.(2019)Stemmler, Ay, Schwenzer, Abel, and
  Bergs}]{Stemmler.2019}
Stemmler, S., Ay, M., Schwenzer, M., Abel, D., and Bergs, T. (2019).
\newblock Model-based predictive force control in milling.
\newblock In \emph{2019 18th European Control Conference (ECC)}, 4313--4318.
  IEEE.
\newblock \doi{10.23919/ECC.2019.8795716}.

\end{thebibliography}
\end{document}